# Fast diffusion of graphene flake on graphene layer

*Irina V. Lebedeva[1,2,3,*], Andrey A. Knizhnik[1,2,†], Andrey M. Popov[4,‡], Olga V. Ershova[3,5], Yurii E. Lozovik[4,3,§], and Boris V. Potapkin[1,2]*

[1]RRC "Kurchatov Institute", Kurchatov Sq. 1, Moscow, 123182, Russia,

[2]Kintech Lab Ltd, Kurchatov Sq. 1, Moscow, 123182, Russia,

[3]Moscow Institute of Physics and Technology, Institutskii pereulok 9, Dolgoprudny, Moscow Region, 141701, Russia,

[4]Institute of Spectroscopy, Fizicheskaya St. 5, Troitsk, Moscow Region, 142190, Russia

[5]School of Chemistry, University of Nottingham, University Park, Nottingham NG7 2RD, United Kingdom

**ABSTRACT**

Diffusion of a graphene flake on a graphene layer is analyzed and a new diffusion mechanism is proposed for the system under consideration. According to this mechanism, rotational transition of the flake from commensurate to incommensurate states takes place with subsequent simultaneous rotation and translational motion until the commensurate state is reached again, and so on. The molecular dynamics simulations and analytic estimates based on *ab initio* and semi-empirical calculations demonstrate that the proposed diffusion mechanism is dominant at temperatures $T \sim (1 \div 3)T_{\text{com}}$, where $T_{\text{com}}$ corresponds to the barrier for transitions of the flake between adjacent energy minima in the commensurate states. For example, for the flake consisting of ~ 40, 200 and 700 atoms the contribution of the proposed diffusion mechanism through rotation of the flake to the incommensurate states exceeds that for diffusion of the flake in the commensurate states by one-two orders of magnitude at temperatures 50 – 150 K, 200 – 600 K and 800 – 2400 K, respectively. The possibility to experimentally measure the barriers to relative motion of graphene layers based on the study of diffusion of a graphene flake is considered. The results obtained are also relevant for understanding of dynamic behavior of polycyclic aromatic molecules on graphene and should be qualitatively valid for a set of commensurate adsorbate-adsorbent systems.

---

[*] lebedeva@kintech.ru

[†] knizhnik@kintech.ru

[‡] am-popov@isan.troitsk.ru

[§] lozovik@isan.troitsk.ru

**PACS** numbers: 65.80.Ck, 68.35.Fx, 68.43.Jk, 85.85.+j

**I. INTRODUCTION**

Since the discovery of graphene, this new material attracted attention of the scientific community due to its unique electronic and mechanical properties[1]. Intensive studies of relative rotational and translational quasistatic motion of graphene layers are currently carried out[2-12]. One of the most interesting phenomenon for graphene goes by the name of "superlubricity", i. e., the ultra low static friction between incommensurate graphene layers[3-12]. Here we for the first time study dynamic behavior of a graphene flake on a graphene layer. Namely, we show that anomalous fast diffusion of the free graphene flake on the underlying graphene layer is possible through rotation of the flake to incommensurate states. As opposed to superlubricity observed in *non-equilibrium* systems (such as a flake moved by the tip of the friction force microscope), diffusion refers to the behavior of a *free* system in *thermodynamic equilibrium*.

The relative motion of flat nanoobjects is determined by the potential energy relief, i.e. the dependence of the interaction energy between the nanoobject and the surface on three coordinates, two of which correspond to the position of the center of mass and the third one is the rotation angle with respect to the surface. The quasistatic superlubricity is observed when the motion takes place across the nearly flat potential energy relief. The fact that this phenomenon is observed for graphene is related to features of the potential energy relief for a graphene flake on a graphene layer. At particular rotation angles, the lattice vectors of the flake can be chosen similar to those of the underlying graphene layer. In these states, the flake is commensurate with the graphene layer and the potential energy relief is highly non-uniform with significant barriers to motion of the flake. At other rotation angles, the flake is incommensurate with the graphene layer and the energy of the flake almost does not depend on its position, i.e. there are no barriers to its motion. Such incommensurate states are observed in the form of so-called Moiré patterns[13,14]. The mechanism of ultra low static friction related to the structural incompatibility of contacting surfaces was first suggested by Hirano and Shinjo[15,16]. Later, Dienwiebel *et al.*[3-5] demonstrated that the static friction force between a graphene flake attached to the tip of the friction force microscope and a graphite surface can become negligibly small at some orientations of the flake. In the present work, we suggest that transition of the flake to the incommensurate states affects not only the *static* tribological behavior of the graphene flake *attached to the tip of the friction force microscope* but also the *dynamic* behavior of the *free* flake and thus might provide anomalous fast diffusion of the flake. Therefore, there is the fundamental distinction between the well-studied superlubricity and the anomalous fast diffusion proposed here in spite of the fact that both phenomena are based on the same features of the potential energy relief.

Incommensurability in adsorbate-adsorbent systems is known to result in fast diffusion of the adsorbate. Experimentally it was demonstrated that large non-epitaxially oriented gold or antimony clusters can diffuse on a graphite surface with a surprising diffusion coefficient of about $10^{-8}$ cm$^2$/s at room temperature[17,18], which is significantly higher than the diffusion coefficients for clusters epitaxially oriented on the surface (of the order of $10^{-17}$ cm$^2$/s, see Ref. [19, 20]). However, such a fast diffusivity was found up to now only for systems where the adsorbate and adsorbent are *not commensurate* at the ground state [21-23]. Based on the systematic study of diffusion mechanisms for a graphene flake on a graphene layer, we suggest that a new diffusion mechanism through rotation of the adsorbate to the "superlubric" incommensurate states (see FIG. 1) is possible and leads to anomalous fast diffusion in adsorbate-adsorbent systems which are interfacially *commensurate* at the ground state.

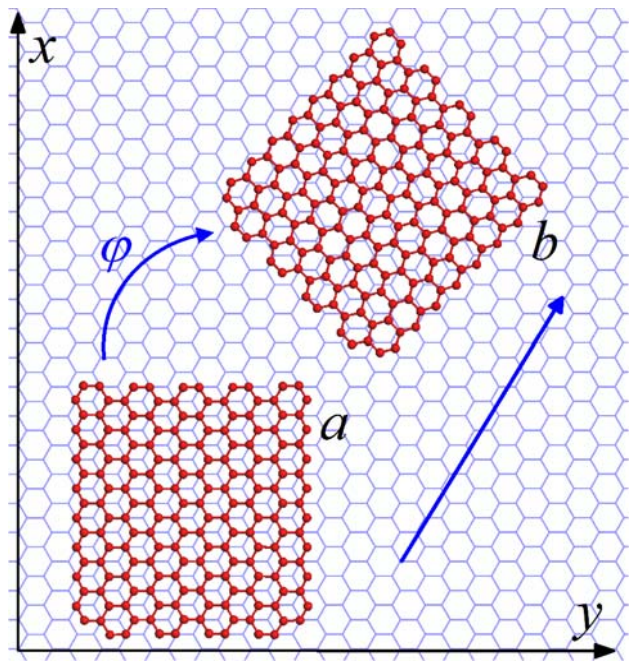

FIG. 1. Schematic representation of the proposed diffusion mechanism through rotation of the flake from the commensurate to incommensurate states. Structure of the graphene flake on the graphene layer: (a) commensurate state, (b) incommensurate state.

In the present work, we perform both density functional theory (DFT) calculations and calculations with empirical potentials to study the dependence of the interlayer interaction energy on the relative position and orientation of graphene layers. These calculations show that the barrier for rotation of the flake to the incommensurate states is of the same order of magnitude as the barrier for motion between adjacent energy minima in the commensurate state. Therefore, there should be a competition between the diffusion mechanisms corresponding the fixed commensurate orientation and incommensurate orientations of the flake. We analyze the diffusion mechanisms both for the incommensurate and commensurate orientations and perform molecular dynamics (MD) simulations of diffusion of a graphene flake on a graphene layer. The estimates and simulations demonstrate that under certain conditions, the proposed diffusion mechanism through rotation of the flake to the incommensurate states is dominant. We believe that a similar diffusion mechanism should also be prominent in any other commensurate adsorbate-adsorbent systems. Particularly, the results obtained here can be useful for understanding of dynamics of polycyclic aromatic molecules on graphene[24].

In addition to the fundamental problems discussed above, the study of diffusion characteristics of graphene is also of interest with regard to the use of graphene in nanoelectromechanical systems (NEMS)[25]. Because of the small size of NEMS, such systems are subject to significant thermodynamic fluctuations[26,27]. On the one hand, relative diffusion[28] or displacement[29] of NEMS components due to thermodynamic fluctuations can disturb the NEMS operation[29,28]. On the other hand, the diffusion can be used in Brownian motors[30]. Experimental measurements of the barriers to relative motion of graphene layers or nanotube walls is also a problem of high importance for elaboration of graphene-based and nanotube-based NEMS[2,31-34]. We propose that experimental measurements of the diffusion coefficient of a graphene flake on a graphene layer can provide the *true* value of the barrier to relative motion of graphene layers.

The paper is organized in the following way. The model used in the calculations and analysis of the dependence of the interlayer interaction energy of graphene layers on their relative position and orientation are presented in Sec. II. Sec. III is devoted to MD simulations demonstrating the proposed diffusion mechanism of a graphene flake. The analytic estimates for the diffusion coefficient at different temperatures and sizes of the flake are obtained in Sec. IV. Our conclusions are summarized in Sec. V.

**II. ANALYSIS OF POTENTIAL ENERGY RELIEF**

To analyze the possible mechanisms of diffusion of a graphene flake on a graphene layer, it is needed to know the potential energy relief of the flake (see FIG. 2). As a model system for energy calculations and MD simulations, we considered a rectangular graphene flake placed on an infinite graphene layer (see FIG. 1). The periodic boundary conditions were applied along mutually perpendicular armchair and zigzag directions to model the infinite substrate layer. In calculations with empirical potentials, the size of the graphene flake was 2.0 nm along the armchair edge and 2.1 nm along the zigzag edge (178 carbon atoms). The size of the model cell was 5.5 nm x 5.7 nm, respectively. The interaction between atoms $i$ and $j$ of the graphene flake and the underlying graphene layer at distance $r_{ij}$ was described by the Lennard–Jones 12–6 potential

$$U_{LJ}(r_{ij}) = 4\varepsilon\left(\left(\frac{\sigma}{r_{ij}}\right)^{12} - \left(\frac{\sigma}{r_{ij}}\right)^{6}\right) \qquad (1)$$

with the parameters $\varepsilon = 3.73$ meV, $\sigma = 3.40$ Å taken from the AMBER database[35] for aromatic carbon. The cut-off distance of the Lennard–Jones potential was taken equal to 20 Å. The Lennard-Jones potential provides the interlayer interaction energy in graphite of about 62 meV/atom, which is consistent with the experimental value 52±5 meV/atom[36]. The covalent carbon-carbon interactions in the layers were described by the empirical Brenner potential[37], which was shown to correctly reproduce the vibrational spectra of carbon nanotubes[38] and graphene nanoribbons[39] and has been widely applied to study carbon systems[26,27,40].

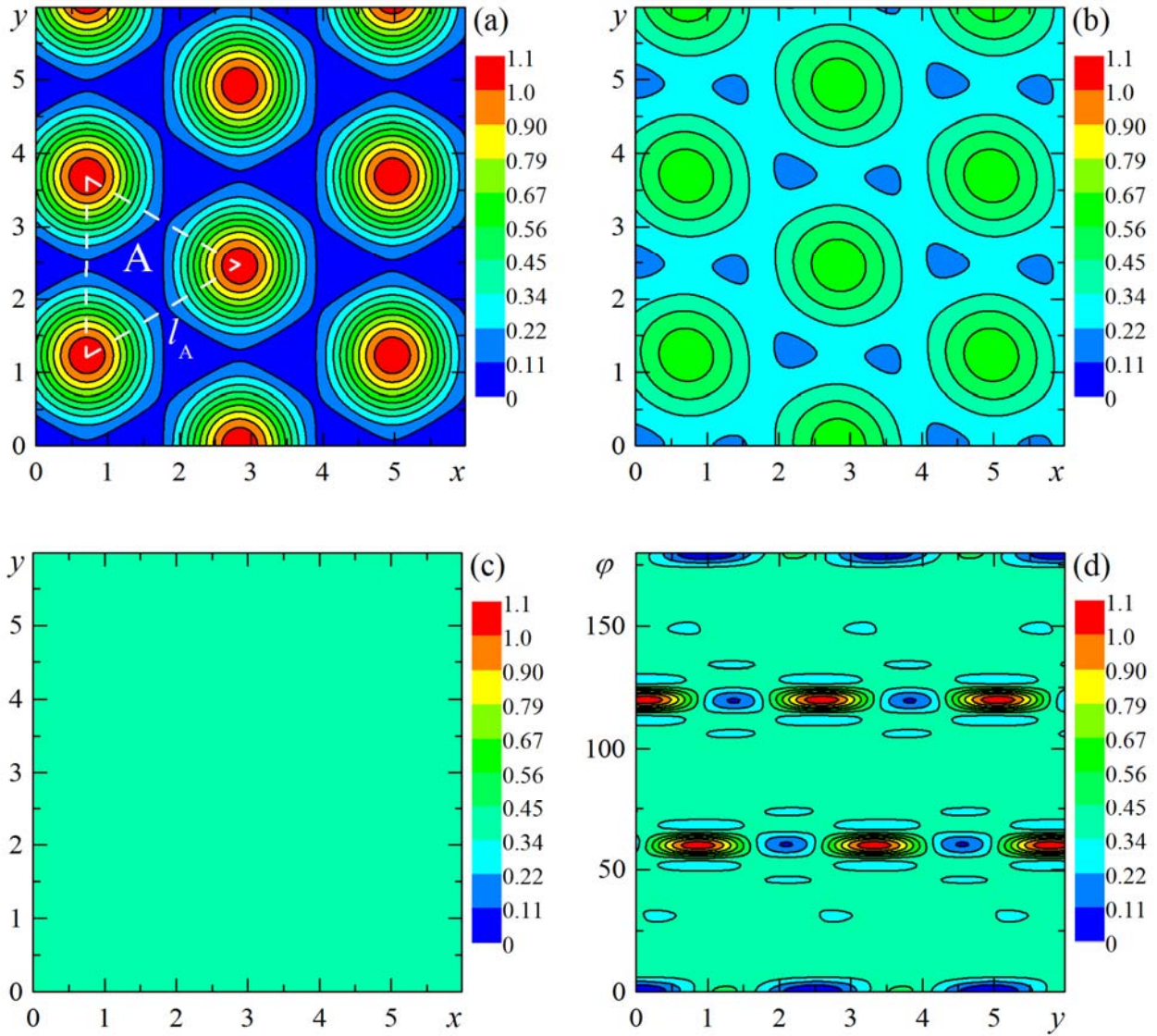

FIG. 2. The interlayer interaction energy between the graphene flake and the graphene layer (in meV/atom) calculated using the Lennard–Jones potential (1) as a function of the position of the center of mass of the flake $x, y$ (in Å, $x$ and $y$ axes are chosen along the armchair and zigzag direction,

respectively) and its relative orientation $\varphi$ (in degrees). (a) $\varphi = 0°$, (b) $\varphi = 4°$, (c) $\varphi = 10°$ and (d) $x = 0$. The energy is given relative to the global energy minimum. Triangle A corresponding to a single energy minimum in the commensurate state is shown with the dashed white lines.

To calculate the dependences of the interlayer interaction energy on the position and orientation of the flake, the structures of the flake and graphene layer were separately relaxed using the Brenner potential and then the flake was rigidly shifted and rotated parallel to the underlying graphene layer. The distance between the flake and the infinite graphene layer was 0.34 nm. The calculated potential energy of the flake as a function of its position and orientation is shown in FIG. 2. The rotation angle $\varphi$ is measured relative to the commensurate orientation of the flake, so that $\varphi = 0°, 60°, 120°$, etc. are attributed to the commensurate states. All other rotation angles correspond to the incommensurate states. The found minimum energy states of the flake correspond to the commensurate AB-stacking, in agreement with the experiment [41]. There is an energy barrier $\varepsilon_{com} = 0.10$ meV/atom for transition of the flake between adjacent energy minima. However, even at temperature above this energy barrier, a long-distance free motion of the flake is not possible due to the numerous energy hills on the potential energy relief, which are higher than the energy barrier $\varepsilon_{com}$ by an order of magnitude. The maximum energy states of the flake correspond to the AA-stacking. The energy difference between the AA and AB-stackings was calculated to be $\varepsilon_{max} = 1.1$ meV/atom.

Based on the approximation[10,42] for the interaction of a single carbon atom in the graphene flake with the graphene layer containing only the first Fourier components, it is easy to show that the potential energy relief for the flake in the commensurate states can be roughly approximated in the form

$$U = U_1 \left( cos(2k_1 x) - 2 cos(k_1 x) cos(k_2 y) \right) + U_0, \qquad (2)$$

where $k_2 = 2\pi/a_0$ and $k_1 = k_2/\sqrt{3}$, $x$ and $y$ axes are chosen along the armchair and zigzag directions, respectively. The parameters $U_1 = 0.225$ meV/atom and $U_0 = -61.92$ meV/atom were fitted to reproduce the potential energy relief calculated using the Lennard–Jones potential. For these parameters, the root-mean square deviation of the potential energy relief (2) from the one calculated using the Lennard–Jones potential equals $0.15 U_1$.

It is seen from FIG. 2 that with rotation of the graphene flake, the magnitude of corrugation of the potential energy relief decreases. At the angle of 10º (see FIG. 2c), the magnitude of this corrugation is negligibly small (less than $0.25 U_1$). At the angle of 60º, the flake becomes again commensurate with the graphene layer. The width of the energy wells and energy peaks in the dependence of the interlayer interaction energy on the orientation of the flake was found to be about $2\delta\varphi \approx 2a_0/L \ll \pi/3$ (see FIG. 2d and FIG. 3), where $a_0 = 2.46$ Å is the lattice constant for graphene and $L$ is the size of the flake, in agreement with Refs. [3-5, 10]. The energy of the incommensurate states relative to the commensurate ones (which is equal to the energy needed to rotate the flake by the angle $\delta\varphi \approx a_0/L$) was calculated to be $\varepsilon_{in} = 0.37$ meV/atom (see FIG. 2 and FIG. 3).

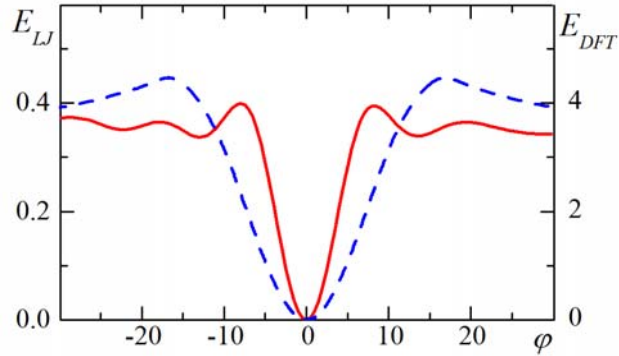

FIG. 3. The interlayer interaction energy between the graphene flake and the graphene layer (in meV/atom) calculated using the Lennard–Jones potential for the 178-atom flake (red solid line; left

axis) and DFT for the 54-atom flake (blue dashed line; right axis) as a function of the rotation angle $\varphi$ (in degrees). The point $\varphi = 0°$ corresponds to the global energy minimum.

Since Lennard-Jones potential was claimed not to be sensitive enough to the relative position of the graphene layers[33], we repeated the similar calculations on the basis of DFT for a smaller system. In these calculations, the graphene flake consisted of 54 carbon atoms and had all edges terminated with hydrogen atoms to prevent distortions of the flake structure at the edges. The size of the model cell was 2.0 nm x 2.1 nm x 1.3 nm. The VASP code[43] with the local density approximation[44] was used. The basis set consisted of plane waves with the maximum kinetic energy of 358 eV. The interaction of valence electrons with atomic cores was described using ultrasoft pseudopotentials[45]. The error in calculations of the system energy was less than 0.005 meV/atom for the chosen cutoff energy of the plane waves. Integration over the Brillouin zone was performed using a single k-point sampling.

According to the DFT calculations, the minimum energy states of the system also correspond to the commensurate AB-stacking. The energy difference between the AA and AB-stackings was calculated to be $\varepsilon_{max} = 10.77$ meV/atom, in agreement with the previous DFT calculations[33]. The barrier for motion of the flake from one energy minimum to another was found to be $\varepsilon_{com} = 1.28$ meV/atom. The relative energy of the incommensurate states was found to be $\varepsilon_{in} = 4.0$ meV/atom (see FIG. 3). The potential energy relief obtained by the DFT calculations was approximated using expression (2) with the parameters $U_1 = 2.38$ meV/atom and $U_0 = -34.04$ meV/atom. The corresponding root-mean square deviation of the potential energy relief (2) equals $0.08 U_1$.

It is seen that the shapes of the potential energy reliefs obtained using the empirical potentials and DFT calculations are qualitatively the same but the magnitudes of corrugation of the interlayer energy differ by an order of magnitude. Therefore, the potential relief of the interaction energy of a graphene flake and a graphene layer can be characterized with a single energy parameter, e.g., $\varepsilon_{\text{com}}$ ($\varepsilon_{\text{in}} \approx 3.5\varepsilon_{\text{com}}$ and $\varepsilon_{\text{max}} \approx 10\varepsilon_{\text{com}}$), which, however, takes different values for different calculation methods. Nevertheless, we show below that the diffusion characteristics of the flake should be mostly determined by the ratio of this energy parameter multiplied by the size of the flake to temperature $\varepsilon_{\text{com}} N / k_{\text{B}} T$ ($k_{\text{B}}$ is the Boltzmann constant and $N$ is the number of atoms in the flake) rather than by the energy parameter alone. Note that the energy parameter $\varepsilon_{\text{com}}$ has not been yet measured experimentally. The results obtained below for certain temperature $T$ and number $N$ of atoms in the flake should be valid for the system with the same ratio $\varepsilon_{\text{com}} N / k_{\text{B}} T$.

Based on the calculations of the potential energy relief for the graphene flake on the infinite graphene layer (see FIG. 2), we propose that different diffusion mechanisms corresponding to commensurate and incommensurate orientations of the flake can be realized depending on temperature and size of the flake. The diffusion of the flake commensurate with the underlying graphene layer at low temperatures $T \ll T_{\text{com}}$, where

$$T_{\text{com}} = N\varepsilon_{\text{com}} / k_{\text{B}}, \qquad (3)$$

can proceed only by rare jumps between adjacent energy minima. On increasing temperature to the region $T_{\text{com}} \le T \ll T_{\text{max}}$, where

$$T_{\text{max}} = N\varepsilon_{\text{max}} / k_{\text{B}}, \qquad (4)$$

the barriers for transitions of the flake between adjacent energy minima become less than the thermal kinetic energy of the flake. So these barriers are not noticeable for the flake during its motion. However, there are still many high potential energy hills, which serve as scattering centers to motion of the flake in the commensurate states and restrict its diffusion length.

Another diffusion mechanism of the flake should be related to rotation of the flake to the incommensurate states. At temperatures $T \ll T_{in}$, where

$$T_{in} = N\varepsilon_{in} / k_B, \tag{5}$$

the probability for the flake to acquire the energy required for rotation to the incommensurate states is small compared to that for jumping between two adjacent energy minima in the commensurate states. However, we show below that this factor can be compensated by long distances passed by the flake before it returns to the commensurate states. Furthermore, on increasing temperature, the time spent by the flake in the incommensurate states also increases. Therefore, we predict that there can be a competition between the diffusion mechanisms for the commensurate and incommensurate states of the flake. At temperatures $T \geq T_{in}$, rotation of the flake becomes almost free, which should provide the dominant contribution of the proposed diffusion mechanism to the diffusion of the flake.

At high temperatures $T \gg T_{max}$, the magnitude of corrugation of the potential energy relief of the graphene flake becomes small compared to the thermal kinetic energy ($k_B T \gg N\varepsilon_{max}$). At these temperatures, the difference between the diffusion of the flake in the commensurate and incommensurate states disappears and the diffusion coefficient of the flake should reach its ultimate value.

## III. MOLECULAR DYNAMICS DEMONSTRATION OF FAST DIFFUSION

To demonstrate that diffusion of a graphene flake can actually proceed through rotation of the flake to the incommensurate states, we performed microcanonical MD simulations of diffusion of the graphene flake at temperatures $T = 50 – 500$ K using the Lennard-Jones potential (1) and the Brenner potential. The flake consisting of 178 atoms was considered. An in-house MD-kMC code was implemented. The time step was 0.4 fs. The initial configuration of the system was optimized at zero temperature. During the simulations, the substrate layer was fixed at three atoms, while the flake was left unconstrained. Transitions of the flake through the boundaries of the model cell were properly taken into account in the calculations of the diffusion length.

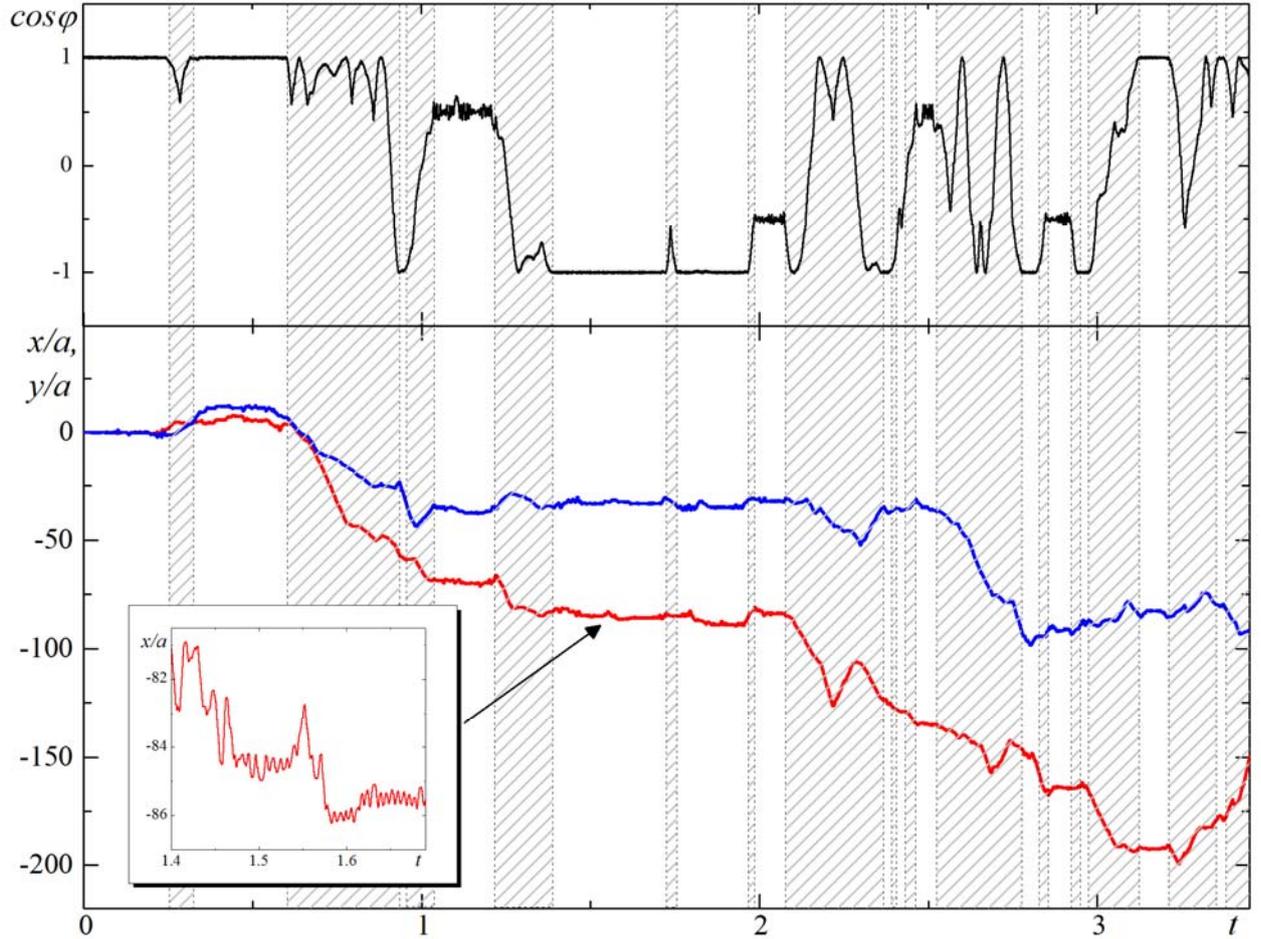

FIG. 4. (Color online) Calculated position $x/a$ and $y/a$ (red and blue lines, respectively; $a = a_0/\sqrt{3}$) of the center of mass of the graphene flake and orientation of the flake $\cos\varphi$ (black line) as functions of time $t$ (in ns) at temperature 300 K. The shaded areas correspond to the fast diffusion of the graphene flake in the incommensurate states. Insert: Scaled-up time dependence of coordinate $x/a$ of the center of mass of the graphene flake corresponding to the slow diffusion of the flake in the commensurate

states.

Let us first discuss the results obtained at room temperature. For the flake consisting of 178 atoms, temperature 300 K corresponds to $T \approx 1.4 T_{com}$ and lies in the range $T_{com} < T < T_{in}$ (see Eqs. (3) and (5)). Therefore, the competition between the diffusion mechanisms in the commensurate states with scattering at the potential energy hills and through rotation of the flake to the incommensurate states should be observed at this temperature. The MD simulations clearly demonstrate that the flake can be found in two states (see FIG. 4). In the first state, the flake stays almost commensurate with the underlying graphene layer and the rotation angle of the flake only slightly oscillates. The transitions of the flake between the energy minima are observed (see the insert of FIG. 4). Nevertheless, the distances passed by the flake across the surface are small and comparable to the lattice constant of graphene. In the second state, as the flake acquires the energy equal to the relative energy of the incommensurate states, it starts the free motion across the surface accompanied by simultaneous rotation.

The trajectories of the flake and the mean-square displacement of the flake obtained on the basis of 6 simulations of 3-3.5 ns duration at temperature 300 K are shown in FIG. 5. It is seen that the asymptotic behavior $\langle r(t)^2 \rangle = 4Dt$, where $D$ is the diffusion coefficient, is almost reached at the times considered. We estimated the diffusion coefficient to be $D = (3.6 \pm 0.5) \cdot 10^{-4}$ cm$^2$/s.

To investigate the behavior of the diffusion coefficient with temperature we performed the MD simulations of the diffusion of the flake in the temperature range of 50 – 500 K. The diffusion coefficients at temperatures 200 and 500 K calculated on the basis of 4 simulations of 1-2 ns duration are given in TABLE I. The same as at 300 K, both diffusion mechanisms for the commensurate and incommensurate states contribute to the diffusion of the flake at these temperatures. At temperatures 50 and 100 K, no rotation to the incommensurate states was detected within the simulation time of a few nanoseconds. Our estimates show that the time required to observe such a rotation at temperatures below 100 K exceeds $0.1\,\mu s$, which is beyond the reach of our MD simulations. Nevertheless, the fact that these events are rare and cannot be observed in the MD simulations does not necessarily mean that they do not provide a noticeable contribution to the diffusion of the flake. So the diffusion coefficients calculated at temperatures 50 and 100 K should be attributed only to the diffusion in the commensurate states and are listed in TABLE II. The contribution of the diffusion mechanism through rotation of the flake to the incommensurate states at these temperatures is estimated in Sec. IV.

TABLE I. Calculated diffusion characteristics of the free flake at different temperatures.

| $T$ (K) | $D$ (cm$^2$/s) | $\langle\tau_{st}\rangle$ (ps) | $\langle\tau_{rot}\rangle$ (ps) | $\langle l^2 \rangle$ (Å$^2$) |
|---|---|---|---|---|
| 200 | $(2.9 \pm 1.7)\cdot 10^{-4}$ | $11 \pm 3$ | $42 \pm 12$ | $300 \pm 100$ |
| 300 | $(3.6 \pm 0.5)\cdot 10^{-4}$ | $13.4 \pm 1.2$ | $20.0 \pm 0.9$ | $240 \pm 30$ |
| 500 | $(7 \pm 3)\cdot 10^{-4}$ | $9.0 \pm 1.1$ | $18.0 \pm 1.5$ | $280 \pm 60$ |

TABLE II. Calculated diffusion characteristics of the flake with the fixed commensurate orientation at different temperatures.

| $T$ (K) | $D_c$ (cm$^2$/s) | $\langle\tau\rangle$ (ps) |
|---|---|---|
| 50 | $(2.6 \pm 0.5)\cdot 10^{-7}$ | $196 \pm 34$ |
| 100 | $(1.09 \pm 0.11)\cdot 10^{-6}$ | $47 \pm 5$ |
| 200 | $(5.3 \pm 0.5)\cdot 10^{-6}$ | $9.5 \pm 0.8$ |
| 300 | $(6.7 \pm 0.6)\cdot 10^{-6}$ | $9.0 \pm 1.0$ |

To clarify the relative contributions of the diffusion mechanisms to the diffusion of the flake, we performed the MD simulations of the diffusion of the flake with the fixed commensurate orientation. The diffusion coefficients $D_c$ calculated on the basis of 2-3 simulations of 1-2 ns duration at temperatures 50 – 300 K are given in TABLE II. From comparison of TABLE I and TABLE II, it follows that at temperatures 200 – 500 K, the diffusion coefficient $D_c$ corresponding to the diffusion of the flake only in the commensurate state is orders of magnitude smaller than the total diffusion coefficient $D$ of the free flake. This proves that the proposed diffusion mechanism through rotation of the flake to the incommensurate states should provide the most significant contribution to the diffusion of the flake under these conditions.

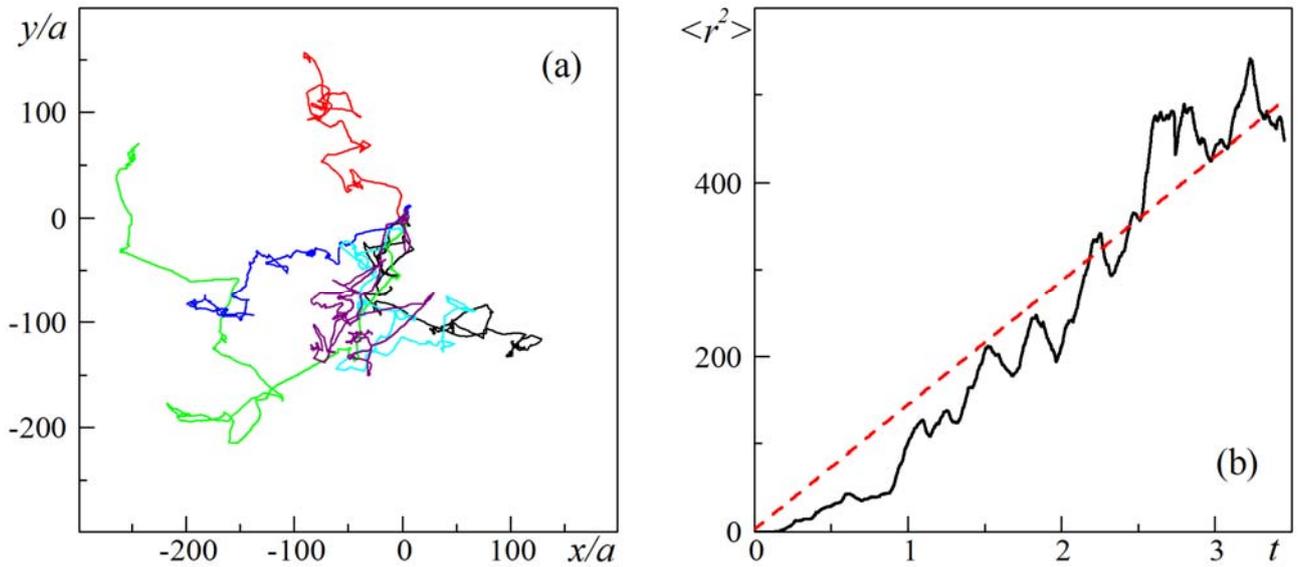

FIG. 5. (a) Calculated trajectories of the flake. (b) Calculated mean-square displacement $\langle r^2(t) \rangle$ (in nm$^2$) of the center of mass of the flake averaged over the trajectories of the flake as a function of time $t$ (in ns) at temperature 300 K (solid line). The dashed line shows the linear approximation of the obtained dependence.

To get a deeper insight into the diffusion mechanisms we performed analysis of the trajectories of the flake (see FIG. 4 and FIG. 5). In the simulations of the diffusion of the free flake, it was assumed that the flake reaches the commensurate state as soon as the rotation angle of the flake gets to the interval $\varphi_0 - \delta\varphi < \varphi < \varphi_0 + \delta\varphi$ ($\varphi_0 = 0°, 60°$, etc.). The average time $\langle \tau_{rot} \rangle$ of rotation by the angle $\Delta\varphi \approx \pi/3 >> 2\delta\varphi$, the average time $\langle \tau_{st} \rangle$ of stay in the commensurate states between these rotations and the mean-square distance $\langle l^2 \rangle$ passed by the flake as it rotates by the angle $\Delta\varphi$ calculated at different temperatures are given in TABLE I. It is seen from TABLE I that in the temperature range of 200 – 500 K, the flake spends most of the time in the incommensurate states ($\langle \tau_{rot} \rangle / \langle \tau_{st} \rangle = 2-4$). This is related to abundance of the incommensurate states compared to the commensurate ones. The average distance $l$ passed by the flake during its rotation by the angle $\Delta\varphi$ considerably exceeds the distance $a = a_0 / \sqrt{3} = 1.42$ Å between adjacent energy minima in the commensurate states (see TABLE I). Further analysis showed that when the flake passes the commensurate state, it can be trapped in a potential energy well (see FIG. 3), reflect from a potential energy hill (see FIG. 2) or continue its rotation in the same direction. Assuming that the flake is trapped in the commensurate state if it stays there for time longer than $\tau'_0 = 6.2$ ps (which corresponds to the period of small rotational vibration in the potential well shown in FIG. 3), the probability for the flake to get trapped in the commensurate states was found to be about 0.27. The probabilities for the flake to reflect from the potential energy hills (see FIG. 2) and to pass the commensurate state were estimated to be about 0.17 and 0.56, respectively. Though the probability for the flake to get trapped in the commensurate states is rather low, the linear velocity of the flake is noticeably changed (by the value $|\Delta\vec{V}| \sim V$, where $V$ is the linear velocity of the flake) almost every time the flake passes the commensurate state, restricting the diffusion

length of the flake. For simplicity, we assume that the translational motion of the flake is strongly disturbed every time as it rotates by the angle $\Delta\varphi \approx \pi/3$ in our estimates in Sec. IV. The simulations for the flake with the fixed commensurate orientation showed that the diffusion of such a flake proceeds by transitions between adjacent energy minima at distance $a = a_0/\sqrt{3} = 1.42$ Å from each other. The average time between these transitions $\langle\tau\rangle$ is given in TABLE II.

**IV. DISCUSSION**

Let us derive an analytic expression for the diffusion coefficient of the flake taking into account the proposed diffusion mechanism. The effective Langevin motion equations for the center of mass and orientation of the flake can be written as

$$M_i \ddot{x}_i(t) = -\frac{\partial U(x_1, x_2, x_3)}{\partial x_i} - \frac{M_i \dot{x}_i(t)}{\tau_i^c} + \xi_i(t), \quad i = 1, 2, 3 \tag{6}$$

where $x_1$, $x_2$ and $x_3$ correspond to the coordinates $x$ and $y$ and angle $\varphi$, respectively; $M_1 = M_2 = M = Nm$ is the mass of the flake; $M_3 = I \propto N^2 m a_0^2$ ($I = 2.7 \cdot 10^{-35}$ g·cm² for the flake under consideration consisting of 178 atoms) is the moment of inertia of the flake; $U(x, y, \varphi)$ is the potential energy relief for the flake on the graphene layer dependent on the two coordinates of the flake and the rotation angle; $\tau_1^c = \tau_2^c = \tau_c$ is the linear velocity correlation time and $\tau_3^c = \tau_c'$ is the angular velocity correlation time characterizing the friction between the flake and the graphene layer; $\xi_i(t)$ is the white noise satisfying the fluctuation-dissipation relation $\langle \xi_i \xi_j \rangle = 2Mk_B T \delta_{ij}/\tau_i^c$.

The characteristics of dynamic friction between the graphene flake and graphene layer were studied using microcanonical MD simulations in the temperature range of 50 – 300 K. The linear and angular velocity correlation times $\tau_c$ and $\tau'_c$ were calculated on the basis of 100 simulations of 10 ps duration at each temperature and were found to be

$$\tau_c^{-1} \approx \tau_c'^{-1} \approx \left(7.5 \cdot 10^{-3} T[\text{K}] + 1.72\right) \cdot 10^9 \, \text{s}^{-1}. \qquad (7)$$

At room temperature, the linear and angular velocity correlation times were found to be approximately $\tau_c \approx \tau'_c \approx 250$ ps.

It is seen from motion equations (6) that while the flake stays in the commensurate state, the translational and rotational motions of the flake are determined by three forces: the potential force $-\partial U / \partial x_i$, the dynamic friction force $-M_i \dot{x}_i / \tau_i^c$ and the random force $\xi_i$. As the flake leaves the commensurate state, the potential force $-\partial U / \partial x_i$ becomes negligibly small compared to the friction and random forces and can be omitted. While the flake rotates by the angle $\Delta\varphi \approx \pi/3$, the rotational and translational motions of the flake can proceed either in ballistic or diffusive regimes depending on the relation between the time $\tau_{\text{rot}}$ of rotation by this angle and the velocity correlation times $\tau'_c \approx \tau_c$. In the ballistic regime ($\tau_{\text{rot}} \ll \tau'_c \approx \tau_c$), the friction between the flake and underlying graphene layer and the thermal noise can be disregarded, i.e. the linear and angular velocities of the flake can be assumed virtually constant. In the diffusive regime ($\tau_{\text{rot}} \gg \tau'_c \approx \tau_c$), the rotational and translational motions of the flake during a single diffusion step are damped and should be described using the diffusion equations with the diffusion coefficients $D_\varphi = k_\text{B} T \tau'_c / I$ and $D_\text{t} = k_\text{B} T \tau_c / M$, respectively, following from the Einstein relation. As soon as the flake reaches again the commensurate state, the potential force becomes relevant. Due to the increased dissipation of the rotational energy and the energy exchange between the rotational and translational degrees of freedom in the commensurate state, the flake can be trapped. However, even if the flake continues its rotation, the potential force disturbs the translational motion of the flake, which limits the diffusion length of the flake in the ballistic regime. In the diffusive regime, the diffusion length of the flake is limited by friction. As shown by the MD simulations, for the flake consisting of about 200 atoms, the ballistic regime is realized in the wide temperature range. The diffusive regime will be considered elsewhere.

Let us consider diffusion of the flake through rotation to the incommensurate states in the ballistic regime ($\tau_{rot} \ll \tau'_c \approx \tau_c$). To estimate the contribution of this diffusion mechanism to the diffusion coefficient of the flake, we assume that the translational motion of the flake is strongly disturbed every time the flake passes the commensurate states and a single diffusion step corresponds to rotation by the angle $\Delta\varphi$, which is supported by the MD simulations. In the ballistic regime ($\tau_{rot} \ll \tau'_c \approx \tau_c$), the angular velocity of the flake does not change significantly within the time of rotation from one commensurate state to another. The time of rotation by a small angle $d\varphi$ at an angular velocity $\omega$ can be found as $d\tau = d\varphi/\omega \cdot f(\omega)d\omega$, where $f(\omega)$ is the probability for the flake in the incommensurate states to rotate with the angular velocity $\omega$. On the other hand, since the average times of stay in the states at an angle $\varphi$ with the angular velocity $\omega$ should have the equilibrium Maxwell–Boltzmann distribution, the time of rotation by the angle $d\varphi$ can also be found as $d\tau \propto exp(-\omega^2/\omega_T^2)d\omega d\varphi$, where $\omega_T = \sqrt{2k_B T/I}$ is the thermal angular velocity of the flake (about $5.5 \cdot 10^{10}$ s$^{-1}$ for the considered flake at room temperature). Based on these two equations, it is seen that $f(\omega) \propto \omega exp(-\omega^2/\omega_T^2)$ and the average time of rotation by the angle $\Delta\varphi \approx \pi/3$ between the commensurate states can be estimated as

$$\langle \tau_{rot} \rangle \approx \frac{2\Delta\varphi}{\omega_T^2} \int_{\omega_{min}}^{\infty} exp\left(-\frac{\omega^2}{\omega_T^2}\right)d\omega \approx \frac{\sqrt{\pi}\Delta\varphi}{\omega_T}, \tag{8}$$

where $\omega_{min} \approx \Delta\varphi / \tau'_c \ll \omega_T$ is the minimum angular velocity for which the condition of ballistic rotation ($\tau_{rot} \ll \tau'_c \approx \tau_c$) is satisfied. At room temperature $\langle \tau_{rot} \rangle \approx 22$ ps, in agreement with the result of the MD simulations (see TABLE I). It is seen that the condition of the ballistic rotation $\langle \tau_{rot} \rangle \ll \tau'_c$ is satisfied for the flake under consideration at room temperature (see Eq. (7)).

In addition to the time $\langle \tau_{rot} \rangle$, which characterizes the time spent by the flake in the incommensurate states in a single diffusion step, the average time of stay in the commensurate states $\langle \tau_{st} \rangle$ between the rotations by the angle $\Delta\varphi$ should be taken into account in calculations of the diffusion coefficient. The average time of stay in the commensurate states can be found from thermodynamic considerations

$$\frac{\langle \tau_{st} \rangle}{\langle \tau_{rot} \rangle} = \phi(T) = \frac{\int_0^{\delta\varphi} \int_{x,y} \exp\left(-\frac{U}{k_B T}\right) dx dy d\varphi}{\int_{\delta\varphi}^{\pi/6} \int_{x,y} \exp\left(-\frac{U}{k_B T}\right) dx dy d\varphi}. \qquad (9)$$

The function $\phi(T)$ was calculated numerically in the wide temperature range. Following the discussion of Sec. II, the function $\phi(T)$ should depend on the ratio of the energy parameter characterizing the potential energy relief of the flake multiplied by the size of the flake to temperature $\varepsilon_{in} N / k_B T \approx 3.5 \varepsilon_{com} N / k_B T$. At temperatures $T < 0.28 T_{in} \approx T_{com}$ (see Eqs. (3) and (5)), the function $\phi(T)$ exponentially decreases with temperature $\phi(T) \approx 0.2 (k_B T / N \varepsilon_{in})^{1.67} \exp(N \varepsilon_{in} / k_B T)$. At temperatures $T > 0.8 T_{in} \approx 2.8 T_{com}$, the temperature dependence of the function $\phi(T)$ is weak and the function reaches about $\phi(T) \approx 0.25$. From Eqs. (8) and (9), it follows that at room temperature $\langle \tau_{st} \rangle \approx 11$ ps, in agreement with the result of the MD simulations (see TABLE I).

Since the linear and angular velocity correlation times are equal $\tau_c \approx \tau'_c$, the velocity of the flake should also be nearly constant during the rotation of the flake by the angle $\Delta\varphi$. Therefore, the diffusion length of the flake corresponding to the rotation by the angle $\Delta\varphi$ can be found as $l = \tau_{\text{rot}} V_T$, where $V_T = \sqrt{2k_B T / M}$ is the thermal linear velocity of the flake ($V_T \approx 4.8 \cdot 10^3$ cm/s at room temperature for the flake consisting of 178 atoms). Analogously to Eq. (8), the mean-square distance passed by the flake while it rotates by the angle $\Delta\varphi$ is given by

$$\langle l^2 \rangle = \frac{2V_T^2 \Delta\varphi^2}{\omega_T^2} \int_{\omega_{\min}}^{\infty} \exp\left(-\frac{\omega^2}{\omega_T^2}\right) \frac{\omega d\omega}{\omega^2} \approx \frac{2V_T^2 \Delta\varphi^2}{\omega_T^2} \ln\left(\frac{\omega_T \tau_c'}{\Delta\varphi}\right). \tag{10}$$

At room temperature, this distance is $l \approx 10a = 1.5$ nm, in agreement with the result of the MD simulations (see TABLE I). This quantity considerably exceeds the distance between adjacent energy minima $a = a_0 / \sqrt{3}$, which corresponds to the diffusion length for the flake with the commensurate orientation.

The contribution of the proposed diffusion mechanism through rotation of the flake to the incommensurate states to the total diffusion coefficient of the free flake can be estimated as

$$D_i \approx \frac{\langle l^2 \rangle}{4(\langle \tau_{\text{st}} \rangle + \langle \tau_{\text{rot}} \rangle)} = \frac{\Delta\varphi}{2\sqrt{\pi}} \frac{V_T^2}{\omega_T (1 + \phi(T))} \ln\left(\frac{\omega_T \tau_c'}{\Delta\varphi}\right). \tag{11}$$

At room temperature, the formula gives $D_i \approx 1.7 \cdot 10^{-4}$ cm$^2$/s, in reasonable agreement with the result of the MD simulations (see TABLE I). The difference by the factor of ~ 2 from the value obtained on the basis of the MD simulations is related to some probability for the flake to skip the commensurate states without disturbing the translational motion of the flake.

Let us discuss the limitations of expression (11). This expression is valid as long as the condition of ballistic rotation $\tau_{\text{rot}} \sim \Delta\varphi / \omega_T \ll \tau_c' \approx \tau_c$ is satisfied and temperature is $T \ll T_{\max}$ (see Eq. (4)). As it follows from Eqs. (7) and (8), the condition of ballistic rotation, $\tau_{\text{rot}} \ll \tau_c' \approx \tau_c$, is violated at very low and high temperatures ($T \leq 10^{-3} T_{\text{com}}$ and $T \gg T_{\text{max}}$ for the flake under consideration). However, the proposed diffusion mechanism through rotation of the flake to the incommensurate states is not dominant in both these cases of very low and high temperatures. At low temperatures $T \ll T_{\text{com}}$, the transition of the flake to the incommensurate states requires the relatively high energy compared to the barrier for transitions of the flake between adjacent energy minima in the commensurate states ($N\varepsilon_{\text{in}} \gg N\varepsilon_{\text{com}}$). Therefore, the diffusion of the flake mostly proceeds in the commensurate states and the proposed diffusion mechanism provides the negligibly small contribution to the diffusion of the flake. At high temperatures $T \gg T_{\text{max}}$, the magnitude of corrugation of the potential energy relief of the graphene flake becomes small compared to the thermal kinetic energy ($k_B T \gg N\varepsilon_{\text{max}}$). In this case, the potential force in motion equations (6) can be disregarded, i.e. the difference between the diffusion of the flake in the commensurate and incommensurate states vanishes. At these temperatures, the diffusion coefficient of the flake reaches its ultimate value determined by friction $D_t = k_B T \tau_c / M = V_T^2 \tau_c / 2$.

The contribution $D_i$ of the proposed diffusion mechanism through rotation of the flake to the incommensurate states to the diffusion coefficient should be compared to the contribution $D_{c1}$ of the diffusion of the flake in the commensurate states. Let us consider diffusion of the flake with the fixed commensurate orientation (such a system was studied by the MD simulations). At temperatures $T \ll T_{\text{max}}$ (see Eq. (4)), the diffusion of the flake in the commensurate states is possible only by transitions between adjacent energy minima. The rate constant for transitions from one energy minimum to another can be found in the framework of the transition state theory (see Ref. [46] and references therein) as

$$k(T) = \frac{3}{2} \frac{\int_{l_A} |V_x| exp\left(-\frac{E}{k_B T}\right) dxdydV_x dV_y}{\int_{S_A} exp\left(-\frac{E}{k_B T}\right) dxdydV_x dV_y} . \quad (12)$$

The expression in the denominator is the integral over the area $S_A$ of the triangle A corresponding to a single energy minimum (shown in FIG. 2). The expression in the numerator is the integral along an edge of the triangle $l_A$. The factor of 1/2 is related to the fact that the motion over the barrier should proceed only in the direction out of the considered energy minimum and the factor of 3 is due to the triangle having three edges. The rate constant $k(T)$ was calculated numerically. Note that the function $k(T)$, analogously to the function $\phi(T)$, depends on the ratio of the energy parameter characterizing the potential energy relief of the flake multiplied by the size of the flake to temperature $\varepsilon_{com} N / k_B T$. Particularly, it can be interpolated as $k(T) \approx 2.1 \tau_0^{-1} exp(-N\varepsilon_{com} / k_B T)$ at temperatures $T < 0.25 T_{com}$ and $k(T) \approx 0.39 \tau_0^{-1} (k_B T / N \varepsilon_{com})^{0.916}$ at temperatures $T > 1.5 T_{com}$, where $\tau_0 \approx 5.8$ ps is the period of small translational vibrations of the flake about the energy minimum.

The average time between transitions of the flake from one energy minimum to another can be found as $\langle \tau \rangle = 1/k$. The diffusion length of the flake in the commensurate state equals the distance $l = a_0 / \sqrt{3}$ between adjacent energy minima. Similar to Eq. (11), the diffusion coefficient of the flake with the fixed commensurate orientation can be found as

$$D_c = \frac{l^2}{4\langle \tau \rangle} = \frac{a_0^2 k(T)}{12} . \quad (13)$$

At room temperature this diffusion coefficient equals $D_c \approx 9 \cdot 10^{-6}$ cm$^2$/s, which is close to the diffusion coefficient estimated on the basis of the MD simulations for the flake in the commensurate state (see

TABLE II and FIG. 6c).

The free flake stays in the commensurate states only for the fraction of time $\alpha_\mathrm{c} = \langle \tau_\mathrm{st} \rangle / ( \langle \tau_\mathrm{rot} \rangle + \langle \tau_\mathrm{st} \rangle ) = \phi/(1+\phi)$, where $\phi$ is defined by Eq. (9). Therefore, the actual contribution of the diffusion of the flake in the commensurate states to the total diffusion coefficient is given by relation $D_\mathrm{c1} = \alpha_\mathrm{c} D_\mathrm{c}$. The fraction $\alpha_\mathrm{c}$ is close to unity at low temperatures ($T \ll T_\mathrm{com}$) and decreases with increasing temperature (see FIG. 6a). This means that the contribution of the diffusion mechanism for the flake in the commensurate states to the total diffusion coefficient is even smaller than $D_\mathrm{c}$. At room temperature for the flake consisting about 200 atoms, the contribution of this diffusion mechanism is only $0.3 D_\mathrm{c}$.

The total diffusion coefficient resulting from the diffusion both in the commensurate and incommensurate states can be found as

$$D = D_\mathrm{i} + D_\mathrm{c1} = D_\mathrm{i} + \frac{\phi}{1+\phi} D_\mathrm{c}. \tag{14}$$

The ratio of the contributions of the diffusion mechanisms $D_\mathrm{i}/D_\mathrm{c1}$ and the dependences of the diffusion coefficients $D$ and $D_\mathrm{c}$ for the free flake and the flake with the fixed commensurate orientation on temperature $T/T_\mathrm{com} = k_\mathrm{B} T / N \varepsilon_\mathrm{com}$ obtained using expressions (11), (13) and (14) are shown in FIG. 6b,c. Let us use FIG. 6 to discuss the diffusion mechanisms of the flake at different temperatures. At low temperatures $T \ll T_\mathrm{com}$, the contributions of different diffusion mechanisms $D_\mathrm{i}$ and $D_\mathrm{c1}$ exponentially depend on the reciprocal of temperature (which is provided by functions $\phi(T)$ and $k(T)$ in expressions (11) and (13)). Since the barrier $N \varepsilon_\mathrm{com}$ for transitions of the flake between adjacent energy minima in the commensurate states is much smaller than the energy $N \varepsilon_\mathrm{in}$ required for rotation of the flake to the incommensurate states $N\varepsilon_{com} \ll N\varepsilon_{in}$, at these temperatures, the flake can stay only in the commensurate states and jump between adjacent energy minima (see FIG. 6a). At $T \sim T_{com}$ the temperature dependences of $D_i$ and $D_{cl}$ switch from the exponential ones to the dependences weaker than linear ones (see Eqs. (11) and (13)). The diffusion mechanism through rotation to the incommensurate states becomes dominant and the ratio of the contributions of the diffusion mechanisms $D_i / D_{cl}$ reaches 10-100 (see FIG. 6b). This is provided both by the decrease of the time spent in the commensurate states (see FIG. 6a) and the long distances passed by the flake in the incommensurate states. As a result, in the temperature range of $T \sim (1 \div 3)T_{com}$, the diffusion coefficient $D$ of the free flake is greater by one-two orders of magnitude than the diffusion coefficient $D_c$ of the flake with the fixed commensurate orientation (see FIG. 6c). This temperature range corresponds to $T \sim 50-150$ K for the size of the flake $N = 40$, $T \sim 200-600$ K for $N = 178$ and $T \sim 800-2400$ K for $N = 700$. At $T \sim T_{in} \approx 3.5 T_{com}$, the flake starts rotating freely. However, the translational motion of the flake is still disturbed as it passes the commensurate states. So the diffusion coefficient is still lowered compared to the maximum diffusion coefficient $D_t = k_B T \tau_c / M$ determined by friction. Only at temperature $T \sim T_{max} \approx 10 T_{com}$, the diffusion coefficient of the flake reaches its ultimate value $D_t$. It is also seen from FIG. 6c that the diffusion coefficients $D$ and $D_c$ estimated on the basis of Eqs. (11) and (13) are in agreement with the results of the MD simulations at different temperatures (see TABLE I and TABLE II).

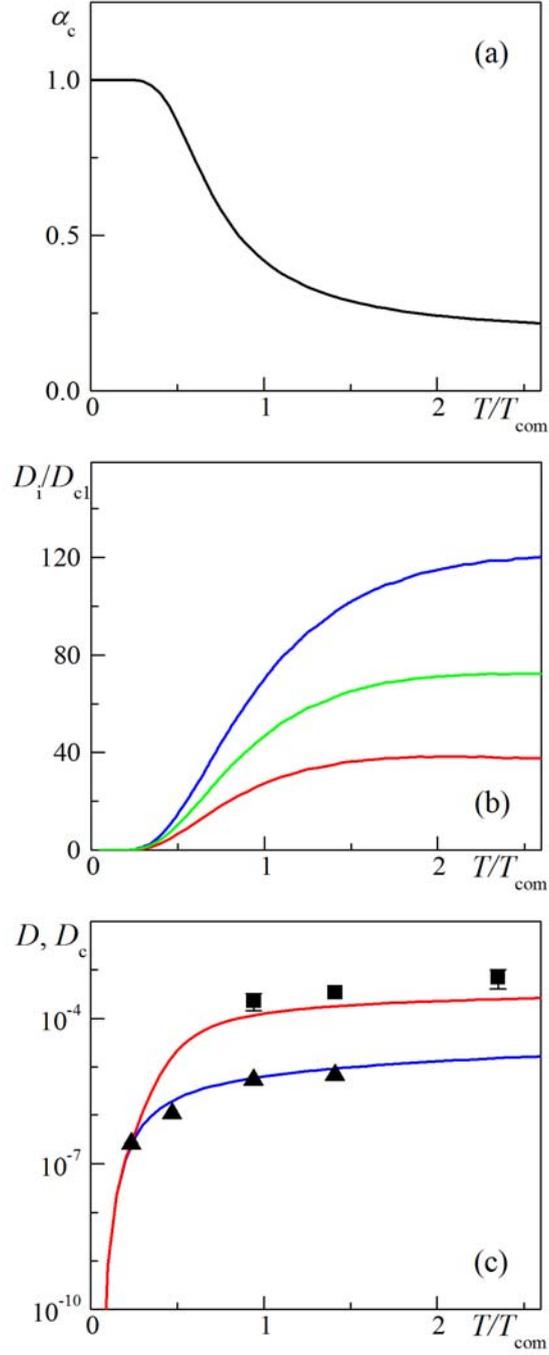

FIG. 6. (a) Calculated fraction of time $\alpha_c$ spent by the flake in the commensurate states as a function of temperature $T/T_{com}$. (b) Calculated ratio of the contributions of the diffusion mechanisms $D_i/D_{cl}$ as a function of temperature $T/T_{com}$ for different sizes of the flake: $N = 40$ (upper blue line), $N = 178$ (middle green line), $N = 700$ (lower red line). (c) Calculated total diffusion coefficient $D$ (red line; in cm$^2$/s) of the free flake and diffusion coefficient $D_c$ (blue line; in cm$^2$/s) of the flake with the fixed commensurate orientation as functions of temperature $T/T_{com}$ for $N = 178$. The results of the MD simulations are shown with black squares for the total diffusion coefficient $D$ of the free flake and with black triangles for the diffusion coefficient $D_c$ of the flake with the fixed commensurate orientation.

As shown above, the magnitudes of corrugation of the potential relief of the interlayer interaction energy between the graphene flake and the graphene layer obtained through the DFT calculations and using the empirical potentials differ by an order of magnitude. However, even using the data obtained by the DFT calculations, the results of our MD simulations and analytic estimates can still be assigned to flakes of smaller size or at higher temperature. From FIG. 6 and Eqs. (11) and (13), it is seen that the diffusion coefficients $D_i$ and $D_{c1}$ depend on the energy parameters of the interlayer interaction in graphite via the factor $T/T_{com} = k_B T / N\varepsilon_{com}$, as discussed above. Therefore, relying on the results of the DFT calculations, it can be, for example, shown that diffusion of a flake consisting of about 70 atoms at room temperature should proceed mostly through its rotation to the incommensurate states ($D_i / D_{c1} \sim 10$).

## V. CONCLUSION

Diffusion of the graphene flake on the graphene layer was analyzed and the new diffusion mechanism through rotation of the flake from the commensurate to incommensurate states was proposed. The molecular dynamics simulations of diffusion of the free flake and the flake with the fixed commensurate orientation were performed in the temperature range of 50–500 K. The analytic expressions for the contributions of the different diffusion mechanisms to the total diffusion coefficient of the flake were obtained. Both the molecular dynamics simulations and estimates based on the analytic expressions demonstrated that the proposed diffusion mechanism is dominant at temperatures $T \sim (1 \div 3)T_{com}$. It was, for example, shown that for the flake consisting of ~ 40, 200 and 700 atoms, the contribution of the proposed diffusion mechanism through rotation of the flake to the incommensurate states exceeds that for diffusion of the flake in the commensurate states by one-two orders of magnitude at temperatures 50 – 150 K, 200 – 600 K and 800 – 2400 K, respectively. We believe that these results can be also applied to polycyclic aromatic molecules on graphene and should be qualitatively valid for a set of commensurate adsorbate-adsorbent systems.

From the analytic expressions derived here, it is seen that the diffusion coefficient of the flake depends exponentially on the difference in the interlayer energies of the commensurate and incommensurate states of the flake and the barrier for transitions of the flake between adjacent energy minima in the commensurate states. Both *ab initio* and semi-empirical calculations were shown to provide similar potential reliefs of the interlayer interaction energy between the graphene flake and the graphene layer (see Eq. (2)), which can be characterized with a single energy parameter. Therefore, we suggest that measurements of the temperature dependence of the diffusion coefficient of a graphene flake on a graphene layer can also give the true value of the barrier for relative motion of graphene layers. In particular, the knowledge of this barrier is valuable for interpretation of the data obtained using the friction force microscope[3-12].

**ACKNOWLEGDEMENT**

This work has been partially supported by the RFBR grants 08-02-00685 and 10-02-90021-Bel.